\input harvmac
\overfullrule=0pt
\parindent 25pt
\tolerance=10000
\sequentialequations


\def\ax{a}

\def\lr{\lref}

\lr\cremmer{E.~Cremmer, B.~Julia and J.~Scherk,
``Supergravity theory in 11 dimensions,''
Phys.\ Lett.\  {\bf B76} (1978) 409.}

\lr\cadavid{A.~C.~Cadavid, A.~Ceresole, R.~D'Auria and S.~Ferrara,
``Eleven-dimensional supergravity compactified on Calabi-Yau threefolds,''
Phys.\ Lett.\  {\bf B357} (1995) 76
[hep-th/9506144].}

\lr\ferrarab{S.~Cecotti, S.~Ferrara and L.~Girardello,
``Geometry Of Type II Superstrings And The Moduli Of Superconformal Field Theories,''
Int.\ J.\ Mod.\ Phys.\  {\bf A4} (1989) 2475.
}

\lr\sabraa{W.~A.~Sabra,
``Black holes in N = 2 supergravity theories and harmonic functions,''
Nucl.\ Phys.\  {\bf B510} (1998) 247
[hep-th/9704147].
}

\lr\sabrab{K.~Behrndt, D.~Lust and W.~A.~Sabra,
``Stationary solutions of N = 2 supergravity,''
Nucl.\ Phys.\  {\bf B510} (1998) 264
[hep-th/9705169].
}

\lr\sabrac{W.~A.~Sabra,
``General static N = 2 black holes,''
Mod.\ Phys.\ Lett.\  {\bf A12} (1997) 2585
[hep-th/9703101].
}

\lr\lusta{K.~Behrndt, I.~Gaida, D.~Lust, S.~Mahapatra and T.~Mohaupt,
``From type IIA black holes to T-dual type IIB D-instantons in N = 2,  D = 4 supergravity,''
Nucl.\ Phys.\  {\bf B508} (1997) 659
[hep-th/9706096].
}

\lr\neugebauer{G.~Neugebauer and D.~Kramer, Ann.\ der Physik (Leibzig) {\bf
24} (1962) 62.}

\lr\gibbonsa{P.~Breitenlohner, D.~Maison and G.~Gibbons,
``Four-Dimensional Black Holes From Kaluza-Klein Theories,''
Commun.\ Math.\ Phys.\  {\bf 120} (1988) 295.
}

\lr\galtsov{G.~Clement and D.~V.~Galtsov,
``Stationary BPS solutions to dilaton-axion gravity,''
Phys.\ Rev.\  {\bf D54} (1996) 6136
[hep-th/9607043].
}

\lr\ferrars{S.~Ferrara and S.~Sabharwal,
``Quaternionic Manifolds For Type II Superstring Vacua Of Calabi-Yau Spaces,''
Nucl.\ Phys.\  {\bf B332} (1990) 317.
}

\lr\ggp{
G.~W.~Gibbons, M.~B.~Green and M.~J.~Perry,
``Instantons and Seven-Branes in Type IIB Superstring Theory,''
Phys.\ Lett.\  {\bf B370} (1996) 37
[hep-th/9511080].
}

\lr\mbgg{
M.~B.~Green and M.~Gutperle,
``Effects of D-instantons,''
Nucl.\ Phys.\  {\bf B498} (1997) 195
[hep-th/9701093].
}

\lr\bbs{K.~Becker, M.~Becker and A.~Strominger,
``Five-branes, membranes and nonperturbative string theory,''
Nucl.\ Phys.\  {\bf B456} (1995) 130
[hep-th/9507158].
}

\lr\bebe{K.~Becker and M.~Becker,
``Instanton action for type II hypermultiplets,''
Nucl.\ Phys.\  {\bf B551} (1999) 102
[hep-th/9901126].
}

\lr\msmg{
M.~Gutperle and M.~Spalinski,
``Supergravity instantons and the universal hypermultiplet,''
JHEP {\bf 0006} (2000) 037
[hep-th/0005068].
}

\lr\stellec{
E.~Cremmer, I.~V.~Lavrinenko, H.~Lu, C.~N.~Pope, K.~S.~Stelle and T.~A.~Tran,
``Euclidean-signature supergravities, dualities and instantons,''
Nucl.\ Phys.\  {\bf B534} (1998) 40
[hep-th/9803259].
}

\lr\bagger{J.~Bagger and E.~Witten,
``Matter Couplings In N=2 Supergravity ,''
Nucl.\ Phys.\  {\bf B222} (1983) 1.
}

\lr\proeyena{B.~de Wit and A.~Van Proeyen,
``Isometries of special manifolds,''
hep-th/9505097.
}

\lr\proeyenb{B.~de Wit, F.~Vanderseypen and A.~Van Proeyen,
``Symmetry structure of special geometries,''
Nucl.\ Phys.\  {\bf B400} (1993) 463
[hep-th/9210068].
}
\lr\ferrarac{I.~Antoniadis, S.~Ferrara, R.~Minasian and K.~S.~Narain,
``R**4 couplings in M- and type II theories on Calabi-Yau spaces,''
Nucl.\ Phys.\  {\bf B507} (1997) 571}

\lr\louisa{H.~Gunther, C.~Herrmann and J.~Louis,
``Quantum corrections in the hypermultiplet moduli space,''
Fortsch.\ Phys.\  {\bf 48} (2000) 119
[hep-th/9901137].}

\lr\stromingera{A.~Strominger,
``Loop corrections to the universal hypermultiplet,''
Phys.\ Lett.\  {\bf B421} (1998) 139
[hep-th/9706195].}

\lr\proeyenc{B.~de Wit and A.~Van Proeyen,
``Symmetries Of Dual Quaternionic Manifolds,''
Phys.\ Lett.\  {\bf B252} (1990) 221.
}

\lr\kallosha{S.~Ferrara, R.~Kallosh and A.~Strominger,
``N=2 extremal black holes,''
Phys.\ Rev.\  {\bf D52} (1995) 5412
[hep-th/9508072].
}

\lr\freb{P.~Fre,
``Supersymmetry and first order equations for extremal states: Monopoles,  hyperinstantons, black holes and p-branes,''
in 
Nucl.\ Phys.\ Proc.\ Suppl.\  {\bf 57} (1997) 52
[hep-th/9701054].
}

\lr\lambert{M.~B.~Green, N.~D.~Lambert, G.~Papadopoulos and P.~K.~Townsend,
``Dyonic p-branes from self-dual (p+1)-branes,''
Phys.\ Lett.\  {\bf B384} (1996) 86
[hep-th/9605146].
}

\lr\denef{F.~Denef,
``Attractors at weak gravity,''
Nucl.\ Phys.\  {\bf B547} (1999) 201
[hep-th/9812049].
}

\lr\stelle{
K.~S.~Stelle,
``BPS branes in supergravity,''
hep-th/9803116.
}

\lr\freh{L.~Andrianopoli, M.~Bertolini, A.~Ceresole, R.~D'Auria, S.~Ferrara, P.~Fre and T.~Magri,
``N = 2 supergravity and N = 2 super Yang-Mills theory on general scalar  manifolds: Symplectic covariance, gaugings and the momentum map,''
J.\ Geom.\ Phys.\  {\bf 23} (1997) 111
[hep-th/9605032].
}

\lr\ceres{A.~Ceresole and G.~Dall'Agata,
``General matter coupled N = 2, D = 5 gauged supergravity,''
hep-th/0004111.
}

\lr\gukov{K.~Behrndt and S.~Gukov,
``Domain walls and superpotentials from M theory on Calabi-Yau  three-folds,''
Nucl.\ Phys.\  {\bf B580} (2000) 225
[hep-th/0001082].
}

\lr\louisb{
K.~Behrndt, C.~Herrmann, J.~Louis and S.~Thomas,
``Domain walls in five dimensional supergravity with non-trivial  hypermultiplets,''
hep-th/0008112.
}

\lr\cecotti{S.~Cecotti, S.~Ferrara and L.~Girardello,
``Geometry Of Type II Superstrings And The Moduli Of Superconformal Field Theories,''
Int.\ J.\ Mod.\ Phys.\  {\bf A4} (1989) 2475.
}

\lr\dewita{J.~De Jaegher, B.~de Wit, B.~Kleijn and S.~Vandoren,
``Special geometry in hypermultiplets,''
Nucl.\ Phys.\  {\bf B514} (1998) 553
[hep-th/9707262].
}

\lr\dewitb{B.~de Wit, B.~Kleijn and S.~Vandoren,
``Superconformal hypermultiplets,''
Nucl.\ Phys.\  {\bf B568} (2000) 475
[hep-th/9909228].
}

\lr\cereb{
A.~Ceresole, R.~D'Auria and S.~Ferrara,
``The Symplectic Structure of N=2 Supergravity and its Central Extension,''
in 
Nucl.\ Phys.\ Proc.\ Suppl.\  {\bf 46} (1996) 67
[hep-th/9509160].
}

\lr\gibbons{G.~W.~Gibbons, M.~B.~Green and M.~J.~Perry,
``Instantons and Seven-Branes in Type IIB Superstring Theory,''
Phys.\ Lett.\  {\bf B370} (1996) 37
[hep-th/9511080].
}

\lr\mooreharvey{
J.~A.~Harvey and G.~Moore,
``Superpotentials and membrane instantons,''
hep-th/9907026.
}

\lr\kachru{S.~Kachru, S.~Katz, A.~Lawrence and J.~McGreevy,
``Open string instantons and superpotentials,''
Phys.\ Rev.\  {\bf D62} (2000) 026001
[hep-th/9912151].
}

\lr\brunner{I.~Brunner and V.~Schomerus,
``On superpotentials for D-branes in Gepner models,''
hep-th/0008194.
}

\lr\lustd{D.~Lust and A.~Miemiec,
``Supersymmetric M5-branes with H-field,''
Phys.\ Lett.\  {\bf B476} (2000) 395
[hep-th/9912065].
}

\lr\minasian{M.~Marino, R.~Minasian, G.~Moore and A.~Strominger,
``Nonlinear instantons from supersymmetric p-branes,''
JHEP {\bf 0001} (2000) 005
[hep-th/9911206].
}

\lr\yujia{M.~Gutperle and Y.~Satoh,
Nucl.\ Phys.\  {\bf B543} (1999) 73
[hep-th/9808080].
}

\lr\yujib{M.~Gutperle and Y.~Satoh,
``D0-branes in Gepner models and N = 2 black holes,''
Nucl.\ Phys.\  {\bf B555} (1999) 477
[hep-th/9902120].
}

\lr\vgg{M.~B.~Green, M.~Gutperle and P.~Vanhove,
``One loop in eleven dimensions,''
Phys.\ Lett.\  {\bf B409} (1997) 177
[hep-th/9706175].
}

\lr\antond{I.~Antoniadis, S.~Ferrara, R.~Minasian and K.~S.~Narain,
``R**4 couplings in M- and type II theories on Calabi-Yau spaces,''
Nucl.\ Phys.\  {\bf B507} (1997) 571
[hep-th/9707013].
}

\lr\aspinwall{P.~S.~Aspinwall,
``Aspects of the hypermultiplet moduli space in string duality,''
JHEP {\bf 9804} (1998) 019
[hep-th/9802194].
}

\baselineskip 20pt plus 2pt minus 2pt
\Title{\vbox{\baselineskip12pt \hbox{hep-th/0010192}
\hbox{HUTP-00/A043}  }}
{\vbox{\centerline{Supergravity Instantons for $N=2$ Hypermultiplets}}}  
\centerline{Michael Gutperle\foot{\tt gutperle@riemann.harvard.edu}
 and Micha\l\ Spali\'nski\foot{\tt mspal@schwinger.harvard.edu}}
\smallskip
\centerline{\it Jefferson Physical Laboratory}
\centerline{\it Harvard University, Cambridge, MA 02138, USA}
\bigskip

\medskip
\centerline{{\bf Abstract}}

The dimensional reduction of eleven dimensional supergravity on a
Calabi-Yau manifold gives $N=2$ supergravity in five dimensions with
$h_{1,1}$ vector and $h_{2,1}+1$ hypermultiplets.  In this paper instanton
solutions are constructed which are responsible for non-perturbative
corrections to the hypermultiplet moduli spaces.  These instantons are
wrapped Euclidean membranes and fivebranes. For vanishing fivebrane charge
the BPS conditions for these solutions define a flow in the hypermultiplet
moduli space and are isomorphic to the attractor equations for four
dimensional black holes.

\noblackbox
\baselineskip 20pt plus 2pt minus 2pt

\Date{October 2000}

\listtoc\writetoc

\newsec{Introduction}

Non-perturbative effects in string theory and M-theory compactifications can
often be described in terms of branes which wrap Euclidean cycles in the
compactification manifold. 
An interesting area 
for 
investigating such effects is $N=2$ supergravity in 
four or five dimensions. 
Membrane and fivebrane instantons will 
provide non-perturbative 
corrections \foot{Some aspects of perturbative corrections to the
universal hypermultiplet obtained by dimensional reduction of higher
derivative terms in M-theory were discussed in
\stromingera\ferrarac\louisa. }
to the metric on the moduli space of the hypermultiplets.
Hypermultiplets of $N=2$ supergravity
parameterize quaternionic manifolds \bagger. The quaternionic geometry and
the relation of the classical hyper and vector multiplet geometries via the
c-map were discussed in \cecotti\ferrars (see also \dewita\dewitb), the
isometries of (special) quaternionic manifolds were discussed in
\proeyena\proeyenb. 
M-brane instanton effects were first discussed by
Becker, Becker and Strominger \bbs. 
The study of such effects was continued in
\bebe, which in particular investigated charge quantization and the
breaking of continuous isometries of the quaternionic manifold due to
instantons and in \msmg\ where supergravity instanton solutions carrying
more than one charge were analyzed.

This paper generalizes the analysis of \bebe\msmg\ , which focused
on the universal hypermultiplet, to an arbitrary number of
hypermultiplets. In a nice paper \lusta, Behrndt et al. used the c-map to
relate four dimensional black hole solutions to four dimensional instanton
solutions, which can be lifted to the five-dimensional solutions of this 
paper. 

The action of the instantons depends on the charges and the values of
complex structure moduli at infinity. Geometrically the action can be
interpreted as the action of a Euclidean membrane wrapping on a
supersymmetric three cycle. It
turns out that for a large class the conditions on BPS-instantons  are 
formally identical to the  attractor equations for $N=2$ black holes
\kallosha\ as formulated by Sabra \sabrac\sabraa. The BPS instanton solution can
be viewed as defining a flow of the hypermultiplet scalars. In the case of
non-vanishing fivebrane charge we find that the 
action has the behavior characteristic of a non threshold bound state,
generalizing the results of \msmg\ obtained for the universal
hypermultiplet.

\newsec{Hypermultiplets in $N=2$ Supergravity}

The supergravity action splits into two parts: one dependent only on vector
multiplets and the other only on hypermultiplets. At the two-derivative
level these two parts are coupled only gravitationally. For the purposes
pursued in this paper only the hypermultiplet part is required. Although
its form is known in the literature, this section will present a derivation
based on compactification of eleven dimensional supergravity on a
Calabi-Yau threefold. This gives the action in five dimensions. The four
dimensional action is essentially the same, since the hypermultiplets are
unchanged by the dimensional reduction. 

The bosonic part of the action of eleven dimensional supergravity
\cremmer\ is given by
\eqn\elevsuga{
S= {1\over 2 k_{11}^2}\int d^{11}x \sqrt{-g}\Big( R-{1\over
48}F^{MNPQ}F_{MNPQ}\big) -{1\over 12k_{11}^2} \int A\wedge F\wedge F \ .  
} 
The supersymmetry transformation of the gravitino in eleven dimensional
supergravity is
\eqn\susytraf{
\delta_\epsilon\psi_M= \partial_M \epsilon +{1\over
4}\omega_{M}^{\underline A\underline B} \Gamma_{\underline A\underline B}
\epsilon -{1\over 288}\Big( \Gamma_{M}^{NPQR}-8\delta_M^N\Gamma^{PQR}\Big)
\epsilon F_{NPQR}\ .
} 
The notation here is that $\underline A,\underline B$ denote tangent space
indices and $M,N$ denote world indices. 
Dimensional reduction of eleven dimensional supergravity on a Calabi-Yau
manifold (with vanishing G-fluxes), produces ungauged five dimensional $N=2$
supergravity with $h_{1,1}$ vector multiplets and $h_{2,1}+1$ 
hyper multiplets \cadavid. The basic 
of special geometry are summarized in appendix A - the conventions used
here will mainly follow \cereb.

The three form field strength of the eleven dimensional supergravity gives
$2h_{2,1}+2$ real scalars $\zeta^I,\tilde\zeta_I, I=0,\cdots,h_{2,1}$.
\eqn\aexans{
C=\sqrt{2}\big( \zeta^I\alpha_I+\tilde\zeta_J\beta^J\big).
}
The ansatz for the eleven dimensional metric is given by
\eqn\metric{
ds^2= e^{-1/3\sigma(y^2)}ds^2_{CY}(z^i,\bar z^i)+ e^{2/3
\sigma(y^2)}g_{\mu\nu}dy^\mu dy^\nu \ ,
}
Here $ds^2_{CY}$ is the Calabi-Yau metric, $\sigma$ parameterizes the
volume of the Calabi-Yau manifold and $z^i$ 
are the complex structure moduli. The K\"ahler moduli are related to vector
multiplets. Since vector multiplets decouple form hypermultiplets
they will not be discussed here. The coordinates of the five
dimensional non-compact space are denoted by $y^\mu$. 

Some details of the dimensional reduction are given in appendices A and
B. The resulting action for $h_{2,1}+1$ hypermultiplets is given by 
\eqn\redac{\eqalign{
S&= {1\over 2 k_5^2}\int d^5x \sqrt{g}\Big\{ {1\over2}
(\partial_\mu \sigma\partial^\mu \sigma) + g_{k\bar l}(z)\partial_\mu 
z^k\partial^\mu z^{\bar l} 
+{1\over 48}e^{-2\sigma}H_{\mu\nu\rho\lambda}H^{\mu\nu\rho\lambda} \cr  
&+{1\over
24}\epsilon_{\mu\nu\rho\lambda\alpha}H^{\mu\nu\rho\lambda}(\zeta^I 
\partial^\alpha \tilde \zeta_I - \tilde \zeta^I
\partial^\alpha \zeta^I)- e^{\sigma} Im{\cal N}_{IJ}\partial_\mu
\zeta^I\partial^\mu\zeta^J\cr
&- e^{\sigma} (Im{\cal N}^{-1})^{IJ}\big(Re{\cal
N}_{IK}\partial_\mu\zeta^K+\partial_\mu\tilde\zeta_I\big)\big(Re{\cal
N}_{JL}\partial_\mu\zeta^L+\partial_\mu\tilde\zeta_J\big)\Big\}}
}
Here $H$ is the field strength of the $3$-form gauge field 
$C_{\mu\nu\rho}$ in five
dimensions arising from the dimensional reduction of $C$, $g$ is the metric
on the complex structure moduli space, and the matrix ${\cal N}$ is
defined in appendix A. Note that  
$Im({\cal N})_{IJ}<0$ hence the  moduli space metric is positive definite
in Minkowski spacetime.  

A three form potential in five dimensions is dual to a (pseudo)scalar
$a$. After dualizing the action takes the form
\eqn\redacb{
\eqalign{S&= \int d^5x \sqrt{g}\Big\{{1\over 2}
(\partial_\mu \sigma\partial^\mu \sigma)+
g_{k\bar l}(z)\partial_\mu z^k\partial^\mu 
z^{\bar l}\cr 
&   +{1\over 2}e^{2\sigma}\big(\partial_\mu \ax+\zeta^I
\partial_\mu \tilde \zeta_I - \tilde \zeta_I
\partial_\mu  \zeta^I\big)^2 - e^{\sigma} Im{\cal
N}_{IJ}\partial_\mu \zeta^I\partial^\mu\zeta^J\cr
&- e^{\sigma} (Im{\cal N}^{-1})^{IJ}\big(Re{\cal
N}_{IK}\partial_\mu\zeta^K+\partial_\mu\tilde\zeta_I\big)\big(Re{\cal
N}_{JL}\partial^\mu\zeta^L+\partial^\mu\tilde\zeta_J\big)\Big\}}
}
Where we set $2k_5^2=1$ for notational convenience.
The hypermultiplet action \redacb\ contains $4(h_{2,1}+1)$ real scalars. Of
these, ($a$, $\sigma$, $\zeta^0$, $\tilde\zeta_0$) comprise the universal
hypermultiplet and ($z^k$, $\bar{z}^{\bar{k}}$, $\zeta^k$, $\tilde\zeta_k$ $k=1,\cdots,h_{2,1}$) 
make up the remaining hypermultiplets. It
is a general consequence of $N=2$ supergravity \bagger\ that these
hypermultiplets parameterize a quaternionic manifold, i.e. $4n$
dimensional manifold with holonomy $Sp(1)\times Sp(n)$.  In \ferrars\
it was shown that the geometry defined by \redacb\ is indeed
quaternionic.

The focus of interest in the following are solutions of the five
dimensional theory which are the analogs of the D-instanton solutions of
ten dimensional type IIB supergravity \gibbons\mbgg. This means that the
solutions satisfy field equations of the Euclidean theory and are localized
in all non-compact five dimensions. Since the metric in the non-compact
dimensions is assumed to be flat and the vector multiplets are decoupled,
one can conclude that the scalars in the vector multiplets must be
constant.

An important consequence of the Euclidean continuation is that the sign of
the ``kinetic'' terms for the scalars $\zeta_I,\tilde\zeta^I, \ax$ in the
action \redacb\ is reversed. This rule for Euclidean continuation of
pseudo-scalars follows from \gibbons\ dualizing the scalars to four forms
(see also \stellec). In the rest of the paper the Euclidean equations are
simply obtained by replacing $\zeta_I\to i\zeta_I,\tilde\zeta^I\to
i\tilde\zeta^I$ in the Minkowskian equations.

\newsec{Isometries and Charges}

The scalars $\zeta^I,\tilde \zeta_I$ and $\ax$ arise from the three 
form potential in eleven dimensions. The dualized action \redacb\ is
invariant under the following shift transformations 
\eqn\isomet{
\zeta_I \to \zeta^I+ \epsilon^I,\quad \tilde\zeta\to
\tilde\zeta_I+\tilde\epsilon_I,\quad  \ax\to \ax+\delta+
\tilde\epsilon^I\zeta_I-\epsilon_I\tilde\zeta^I .
}
The  currents associated with these shifts of are
\eqn\currents{
\eqalign{j^I_\mu&=e^{\sigma}\Big(- Im{\cal N}_{IJ}\partial
\zeta^J-Re {\cal N}_{IJ} (Im{\cal N}^{-1})^{ JL}(Re {\cal
N}_{LM}\partial_\mu 
\zeta^M+\partial_\mu \tilde{\zeta}_L)\Big) \cr
&\;+ e^{2\sigma}
(
\partial_\mu\ax+\zeta^I\partial_\mu\tilde\zeta_I-\tilde\zeta^I\partial_\mu
\zeta_I)\tilde \zeta^I , \cr
\tilde j^I_\mu&=-2 e^{\sigma} (Im{\cal N}^{-1})^{ IL}(Re {\cal
N}_{LM}\partial_\mu 
\zeta^M+\partial_\mu \tilde{\zeta}_L)-e^{2\sigma}
(
\partial_\mu\ax+\zeta^I\partial_\mu\tilde\zeta_I-\tilde\zeta^I\partial_\mu
\zeta_I)\zeta^I , \cr
j^5_\mu &= e^{2\sigma} \big(\partial_\mu \ax+\zeta^I
    \partial_\mu \tilde \zeta_I - \tilde \zeta^I
    \partial_r  \zeta_I\big).
}}
The corresponding charges are
\eqn\charges{
Q^I= \oint d\Sigma^\mu j^I_\mu,\quad
\tilde Q^I= \oint d\Sigma^\mu\tilde j^I_\mu ,\quad
Q^5 = \oint d\Sigma^\mu j^5_\mu .
}
Note that the charges $Q^I$, $\tilde Q^I$ are not invariant under the
shifts; they transform as
\eqn\shiftch{
Q_5\to Q_5,\quad Q^I\to Q^I+\tilde\epsilon^I Q_5,\quad \tilde
Q^I\to \tilde 
Q^I- \epsilon^I Q_5 .
}
In general there will be additional isometries of the hypermultiplet moduli
space which depend on the details of the geometry. For example, for
quadratic prepotentials $F=1/2 \sum_{I=0}^n(Z^I)^2$ (which however do  not
arise in Calabi-Yau 
compactifications) the geometry is the coset manifold
$SU(2,n+1)/(U(2)\times SU(n+1))$ which the isometry group $SU(2,n+1)$.  A
detailed 
discussion of symmetries of special quaternionic manifolds can be found
in \proeyena\proeyenb \proeyenc.

\newsec{Equations of Motion}

The equations of motion for $\zeta^I$, $\tilde \zeta_I$ and $\ax$ 
are simply the conservation equations for the associated currents
\eqn\eqofmha{
\eqalign{
\partial_\mu j^{mI}&=0,\quad \partial_\mu \tilde
j^{mI}=0,\quad I=0, \cdots, h_{2,1} ; \cr
\partial_\mu j^{5 m}&=0 .
}}
The $\sigma$ equation of motion is given by (note that we display the
Euclidean equations of motion)
\eqn\eqofmhb{\eqalign{
-\partial^2\sigma -  e^{2\sigma}\big(
\partial_\mu \ax+\zeta^I \partial_\mu \tilde \zeta_I -\tilde \zeta^I
\partial_r \zeta_I\big)^2 + e^{\sigma} Im{\cal 
N}_{IJ}\partial_\mu \zeta^I\partial^\mu\zeta^J\cr
+ e^{\sigma} (Im{\cal N}^{-1})^{IJ}\big(Re{\cal
N}_{IK}\partial_\mu\zeta^K+\partial_\mu\tilde\zeta_I\big)\big(Re{\cal
N}_{JL}\partial^\mu\zeta^L+\partial^\mu\tilde\zeta_J\big)=0 .
}}
The equation of motion for the scalar $z^k$ is 
\eqn\eqofmc{\eqalign{
\partial^2 z^k + \Gamma^k_{lm} \partial_\mu z^l\partial^\mu z^m+{1\over
2}e^{\sigma} 
g^{k\bar l} \partial_{\bar l} \Big\{ Im{\cal
N}_{IJ}\partial_\mu 
    \zeta^I\partial^\mu\zeta^J\cr
+(Im{\cal N}^{-1})^{IJ}\big(Re{\cal
N}_{IK}\partial_\mu\zeta^K+\partial_\mu\tilde\zeta_I\big)\big(Re{\cal
N}_{JL}\partial^\mu\zeta^L+\partial^\mu\tilde\zeta_J\big)\Big\}=0 .
}}
The condition $R_{\mu\nu}=0$ implies the vanishing of the corresponding
components of the energy-momentum tensor, i.e.
\eqn\vanact{
\eqalign{
& {1\over 2}\partial_\mu \sigma \partial_\nu \sigma +
g_{k\bar l}(z)\partial_\mu z^k\partial_\nu z^{\bar l}
+ e^{\sigma} Im{\cal N}_{IJ}\partial_\mu
\zeta^I\partial_\nu\zeta^J\cr 
&-{1\over 2}e^{2\sigma}
\big(\partial_\mu \ax+\zeta^I\partial_\mu \tilde \zeta_I - 
\tilde \zeta^I \partial_\mu  \zeta_I\big)
\big(\partial_\nu \ax+\zeta^I\partial_\nu \tilde \zeta_I - 
\tilde \zeta^I \partial_\nu  \zeta_I\big)\cr
&+ e^{\sigma} (Im{\cal N}^{-1})^{IJ}\big(Re{\cal
N}_{IK}\partial_\mu\zeta^K+\partial_\mu\tilde\zeta_I\big)\big(Re{\cal
N}_{JL}\partial_\nu\zeta_L+\partial_\nu\tilde\zeta_J\big)=0 .
}}
Note that this condition implies the vanishing of the bulk part of the
action for the Euclidean instanton solutions.

\newsec{A Simple Solution}

A simple solution is given by taking the $z^k$ to be constant and real (and
as a consequence ${\cal N}_{IJ}$ to be purely imaginary and negative
definite). The fields $\zeta^I$ and $\ax$ can consistently be set to
zero. The relevant part of the action is then given by
\eqn\relaca{
S=\int d^5x\sqrt{g}\big( {1\over 2}(\partial\sigma)^2 -
e^{\sigma} (Im{\cal 
N}^{-1})^{IJ}\partial \tilde \zeta_I \partial \tilde \zeta_J\big) .
}  
Passing to Euclidean space yields 
\eqn\relaca{
S=\int d^5x\sqrt{g}\big({1\over 2} (\partial\sigma)^2 + 
e^{\sigma} (Im{\cal 
N}^{-1})^{IJ}\partial \tilde \zeta_I \partial \tilde \zeta_J\big) .
} 
The instanton solutions are assumed to be spherically symmetric in the five
Euclidean dimensions. Specifically, the following ansatz is made:
\eqn\solsimpl{
\sigma = 2\ln h,\quad \tilde\zeta_I = \alpha_I {1\over h}+const .
}
The function $h$ is harmonic in five dimensions
\eqn\harmf{
h(r)=e^{\sigma_\infty/2} + { q\over {3 r^3}} .
}
The ansatz \solsimpl\ solves the Euclidean equations of motion provided the 
constant vector $\alpha_I$ satisfies $(Im{\cal
N}^{-1})^{IJ}\alpha_I\alpha_J=-2$. One can calculate the non-vanishing
charges carried by this solution:
\eqn\charofsol{
\tilde Q^I = - Vol(S^4) q (Im{\cal N}^{-1})^{IJ}\alpha_J .
}
This leads to a relation between the quantity $q$ in \harmf\ and the
charges $\tilde Q^I$: 
\eqn\chasimpl{
q= \alpha_I\oint d\Sigma^\mu \tilde j_\mu^I .
}
The bulk part of the action for the instanton vanishes as discussed in
\ggp\mbgg\ , and the non-zero contribution comes solely from a boundary
term:   
\eqn\sinsta{
S_{inst}= - \oint d\Sigma^\mu \partial_\mu \sigma = 
e^{-\sigma_\infty/2} q . 
}
These solutions are the simplest generalizations of the single charged
instanton solutions for the universal hypermultiplet found in
\bbs\bebe\msmg\ and correspond to Euclidean M2 
branes wrapped on a three-cycle in the CY (the charge vector $\alpha_I$ is  
related to the homology of the three-cycle). This solution is in fact a BPS
solution which preserves sixteen of the thirty two supersymmetries, as will
be discussed in section 9.

\newsec{Instanton Equations, Harmonic Functions and Geodesics}

When the vector multiplets are neglected the bosonic $N=2$ action is given
by a quaternionic sigma model of the hypermultiplets couples to gravity.
The action is has the form 
\eqn\acsigm{
S= \int d^5x\sqrt{g}\big(R-{1\over 2} {\cal G}_{uv}(\phi)\partial_\mu
\phi^u\partial^\mu\phi^v\big),
}
where $u,v=1,\cdots, 4(h_{2,1}+1)$, so that the $\phi^u$ now denote all the
hypermultiplet (pseudo)scalars and ${\cal G}$ is the quaternionic metric
determined by the action \redacb .

The solutions of interest here have the property that the five dimensional
metric is flat, which implies that
\eqn\flatm{
R_{\mu\nu}= \half g_{uv}\partial_\mu\phi^u\partial^\mu \phi^v =0 . 
}
A simple ansatz for finding such solutions was presented in
\neugebauer\gibbonsa\galtsov\ : the dependence of the scalar fields
$\phi^u$ on the spacetime coordinates $x^\mu$ is 
through a scalar function $\sigma(x)$, i.e. $\phi^u(x)=\phi^u(\sigma(x))$.
The equation of motion for $\phi$ then becomes
\eqn\eqofmph{
\nabla^2 \sigma (\phi^u)^\prime + \partial_\mu
\sigma\partial^\mu\sigma\big[(\phi^u)^{\prime\prime} +
\Gamma^u_{vw}(\phi^v)^\prime(\phi^w)^\prime\big]=0  ,
} 
where $(\phi^k)^\prime=\partial_\sigma \phi^k$.
The first term in this equation is the spacetime Laplace operator acting on
the scalar function $\sigma(x)$. Hence if $\sigma(x)$ is a harmonic
function in spacetime  the
remaining part of \eqofmph\ is nothing but the geodesic
equation in the moduli space where the scalar $\sigma$ is now interpreted
an affine parameter. 
Since one is seeking a solution where the spacetime is flat, the
gravitational 
part of the equations of motion \flatm\ implies $R_{\mu\nu}=0$  which means
that the geodesic is null.

This construction can be generalized by allowing the fields $\phi^k$ 
to depend on several harmonic maps $\sigma^a, a=1,\dots, n$. A solution to
the equations of motion is then given by a totally geodesic null
submanifold. Supersymmetry conditions impose further constraints on the
solutions. For general moduli spaces finding such geodesic submanifolds is
very complicated. However if the moduli space is a coset manifold one can
apply the construction given in \galtsov\ to construct the solutions. 

In the case of a single (universal) hypermultiplet, the moduli space is
given by the coset $SU(2,1)/U(2)$, in toroidally compactified type II
theories the scalar moduli space is given the well known cosets $G/H$,
instanton solutions in these theories were discussed in \stellec.  
Representing the elements of the coset $G/H$ as matrices $g$ the equation
of motion for the scalars becomes
\eqn\scalareqm{
\partial^\mu(g^{-1}\partial_\mu g)=0 .
}
The matrix $g$ can be parameterized by \galtsov\stelle\
\eqn\matrip{
g= a\exp(\sum_i b_i \sigma_i) ,
}
where $\sigma_i$ are harmonic function in the space time. The conditions
that $\sigma_i$ parameterize a null geodesic submanifold translate into
certain conditions on the matrices $b_i$ (see \galtsov\stelle\ for
details). Hence finding the instanton solutions for hypermultiplets which
are coset spaces can be reduced to solving algebraic matrix equations.
However the hypermultiplet geometry for Calabi-Yau compactifications is in
general not a coset manifold (since the associated prepotential is not
quadratic). The techniques reviewed above might however still be useful if
one can find a subspace of the full hypermultiplet moduli space which 
is a coset.
In addition supersymmetry does not make a direct appearance in this
discussion, yet as will become clear it provides further constraints on the 
velocity vectors of the geodesics. Therefore the general formalism might
also be used to find non-supersymmetric solutions.

\newsec{Supersymmetry}

The supersymmetry transformations for the fermionic hyperino and gravitino
fields can be derived by dimensional reduction of the eleven
dimensional supersymmetry transformation rules \susytraf. Some details of
this can be found in appendix B. The
transformation law of the gravitino is given by 
\eqn\susygrav{
\delta \psi^A_\mu =(\partial_\mu+{1\over
4}\omega_\mu^{ab}\gamma_{ab})\epsilon^A+(Q_\mu)_B^{\;A}\epsilon^B ,
}
where $(Q_\mu)_B^{\;A}$ is a composite $Sp(1)$ gauge connection 
defined by
\eqn\qdeft{
(Q_\mu)_B^{\;A}=\pmatrix{{1\over 4}\big(v_\mu-\bar v_\mu -{\bar ZN
\partial_\mu Z-ZN\partial_\mu\bar Z\over \bar ZNZ}\big)& -\bar u_\mu\cr
u_\mu & 
-{1\over 4}\big(v_\mu-\bar v_\mu -{\bar ZN 
\partial_\mu Z-ZN\partial_\mu\bar Z\over \bar ZNZ}\big) } .
}
The supersymmetry transformations of the hyperinos are 
\eqn\susyhyp{\eqalign{
\delta\xi^I_1&=e^{1I}_\mu\gamma^\mu\epsilon_1 -\bar
e^{2I}_\mu\gamma^\mu\epsilon_2\cr 
\delta\xi_2^I&=e^{2I}_\mu \gamma^\mu \epsilon_1+\bar
e^{1I}_\mu\gamma^\mu\epsilon_2 .
}} 
The vielbein components $e^{I1}_\mu,e^{I2}_\mu$ are defined in terms of the 
scalar fields in the following way
\eqn\vielbcomp{
e^{1\;I}_\mu=\pmatrix{u_\mu\cr E^A_\mu},\quad
e^{2\;I}_\mu=\pmatrix{v_\mu\cr e^A_\mu}}
where
\eqn\vielba{\eqalign{u_\mu&= e^{1/2\sigma}L^I \big( {\cal
N}_{IJ}\partial_\mu\zeta^J+\partial_\mu\tilde\zeta_J\big)\cr
v_\mu&=  {1\over 2}\partial_\mu\sigma+ {i\over 2} e^{\sigma}\big
( \partial_\mu\ax+\zeta^I\partial_\mu\tilde\zeta_I-\tilde\zeta^I\partial_\mu \zeta_I\big)\cr
e^A_\mu&= e^A_i\partial_\mu z^i\cr
E^A_\mu&= e^{\sigma/2}e^{A i}  f_{ i}^I \big( \bar{\cal
N}_{IJ}\partial_\mu\zeta^J+\partial_\mu\tilde\zeta_I\big)
}}
Here  $e^A_i$ is the vielbein associated with the metric on the moduli
$z^i$, i.e. $\delta_{AB}e^A_i e^B_{\bar j}= g_{i\bar j}$. This form of the
components 
agrees with the ones given in \ferrars. 
The components \vielbcomp\ can be combined into a quaternionic vielbein
\eqn\quatvb{
V^{\alpha A}= \pmatrix{e_1^I\cr \bar e_2^I \cr -e_2^I\cr \bar
e_1^I},\quad\quad  \alpha=1,\cdots,2 h_{2,1},\; A=1,2.
}
The vielbein satisfies the reality constraint $(V^{\alpha A})^*=
\epsilon_{AB}C_{\alpha\beta}V^{\beta B}$  (this shows the connection with
the notation used in  \freb\ceres). 

The action \redacb\ can be expressed in the following way
\eqn\acwvb{\eqalign{
S&= 2 \int d^5x \sqrt{g}\Big\{ \sum_I
( e^{1I}_\mu\bar e^{1I}_\mu+ e^{2I}_\mu \bar e^{2I}_\mu)\}\cr
&=2 \int d^5x \sqrt{g}\Big\{ 
u_\mu\bar u_\mu+ v_\mu \bar v_\mu + \sum_A(e^A_\mu \bar e^A_\mu+E^A_\mu
\bar E^A_\mu)\} ,
}}
where the following useful identity of special geometry  was employed 
\eqn\identuse{
f_i^I g^{i\bar j} f_{\bar j}^J =- {1\over 2} (Im{\cal
N}^{-1})^{IJ}- e^K \bar Z^I Z^J .
}
Note that this formula shows that $Im {\cal N}$ is negative definite since 
the Minkowskian metric has to be positive definite.

For a scalar field configuration that preserves half of 
the supersymmetries 
the  variations $\delta\xi^k=0$ have to vanish, for a pair of
spinors $\epsilon^A$. This condition can be
interpreted as the vanishing of a velocity vector defined by \susyhyp\ for
suitably chosen $\epsilon^A$.  
For rotationally invariant field configurations the hyperino
transformations  can
be interpreted as  $2\times2$ matrix equations and the BPS condition
is that the determinant of these matrices
vanishes
\eqn\detcond{
\epsilon^{1I}_\mu\bar\epsilon^{1I}_\mu +
\epsilon^{2I}_\mu\bar\epsilon^{2I}_\mu=0,\quad I=0,\cdots, h_{2,1}.
}
This condition can be interpreted as follows: As discussed in section 7
the instanton solution is given by a geodesic submanifold, parameterized by
a number of harmonic functions.  The BPS condition implies that the
submanifold is 
'null' with respect to all the inner products defined in  \detcond. Note
that it follows that the solution satisfies   \vanact\ since this is simply
the sum of \detcond\ for $I=0,\cdots,h_{2,1}$.

\newsec{BPS Solutions and Attractor Equations}

It is straightforward to check that for the simple solution carrying
one charge described in section 6, only the $\delta\xi^0$ variation is
nontrivial 
and that the solution preserves half the supersymmetries. 
It is also possible to write down solutions which carry arbitrary charges
$Q^I$ and $\tilde Q^I$. For the moment however the five brane charge $Q_5$
will be set to zero. This is equivalent to  imposing
\eqn\fbcvan{
\partial_\mu \tilde \phi +\zeta^I\partial_\mu
\tilde\zeta_I-\tilde\zeta_I\partial_\mu\zeta^I=0 .
}  
A family of solutions depending on $2(h_{2,1}+1)$ harmonic functions can be 
constructed. Let
\eqn\harmfct{
H_I = h_I+ {q_I\over {3 r^3}},\quad 
\tilde H^I= \tilde h^I+{\tilde q^I\over {3 r^3}} .
}
In the solution is scalars $\zeta^I,\tilde \zeta_I$ are taken to satisfy
the ansatz
\eqn\zetdef{
\eqalign{\partial_\mu \zeta^I &= -e^{-\sigma} \Big\{(Im{\cal
N}^{-1})^{IJ}\partial_\mu H_J -(Im{\cal
N}^{-1})^{IJ}(Re{\cal N})_{JK}\partial_\mu \tilde H^K\Big\},\cr
\partial_\mu \tilde \zeta _I&= -e^{-\sigma} \Big\{(Im{\cal
N})_{IJ}\partial_\mu \tilde H^J -(Re{\cal N})_{IJ}(Im{\cal
N}^{-1})^{JK}\partial_\mu  H_K\cr
&+ (Re{\cal N })_{IJ}{Im \cal
N}^{-1})^{JK}(Re{\cal N })_{KL}\partial_\mu \tilde H^L\Big\} .
}}
With this it is easy to see that the charges defined in the currents
\currents\ for this solution are given by
\eqn\chargeso{
Q_5=0,\quad 
Q^I= Vol(S^4) q^I,\quad 
\tilde Q^I= Vol(S^4) \tilde q^I .
}
The BPS condition following from the vanishing of the hyperino variation
can be written as (using \susyhyp\ and \vielba)
\eqn\susyvhp{
\eqalign{{1\over 2}\partial_\mu \sigma \gamma^\mu \epsilon_2-
e^{\sigma/2}L^I({\cal N}_{IJ}\partial_\mu \zeta^J+\partial_\mu
\tilde\zeta_I)\gamma^\mu\epsilon_1&=0, \cr 
\partial_\mu z^i\gamma^\mu \epsilon_1 + e^{\sigma/2} g^{i\bar j}f_{\bar j}^I({\cal
N}_{IJ}\partial_\mu \zeta^J+\partial_\mu 
\tilde\zeta_I)\gamma^\mu\epsilon_2&=0 .
}} 
After continuation to Euclidean space the BPS equations for rotationally
symmetric field configurations can be derived by choosing the spinors
$\epsilon_1=\pm\epsilon_2$. Using \zetdef\ one finds that \susyhyp\ turns
into 
\eqn\bpsflob{\eqalign{
{1\over 2}{d \sigma\over dr} - L^I(q_I-{\cal N}_{IJ}\tilde q^J) 
{e^{-\sigma/2}\over r^4}&=0, \cr
{d z^i\over dr}+ g^{i\bar j}\bar f_{\bar j}^I(q_I-{\cal N}_{IJ}\tilde q^J)
{e^{-\sigma/2}\over r^4}&=0 .
}} 
The equations \bpsflob\ are exactly of the same form as the 
attractor equations for $N=2$ black holes \kallosha\ as written\foot{This
involves identifying $-\sigma/2$ with Sabra's $U$ and the charges $q^I,\tilde
q^I$ with dyonic charges of the black hole solution.} in
\freb\sabrac\sabraa. 

The solution is further specified by expressing the scalar fields $z^i$ 
in terms of the harmonic functions \harmfct\ 
\eqn\sabr{
i(Z^I-\bar Z^I)= \tilde H^I,\quad i(F_I-\bar F_I)= H_I,
}
and the volume scalar $\sigma$ is given by
\eqn\volsc{
\sigma = \ln\;i(\bar Z^IF_I- Z^I\bar F_I).
}
There is an additional constraint on the harmonic functions 
\eqn\constrA{
H_I \partial_\mu \tilde H^I- \tilde H^I\partial_\mu H_I=0
}
which guarantees the integrability of the condition \fbcvan\ for the 
vanishing fivebrane charge.

In \sabraa\ Sabra has  shown that the ansatz \sabr,\volsc\ solves
the attractor equations \bpsflob. From this it follows that the our ansatz
defines a BPS configuration. Sabra's calculation will not be repeated
here. 
Instead, in appendix C it is demonstrated that the solution also satisfies 
the Euclidean equations of motion as expected. 

Note that the attractor equation \bpsflob\ together with \zetdef\ defines a
flow of the hypermultiplet scalars when $r$ varies from to $r=\infty$ to
$r=0$ for an instanton solutions with given charges $q_I,\tilde q^I$.

\newsec{The Instanton Action}

As discussed in section 8 the BPS equations \susyvhp\ imply the vanishing
of the bulk part of the action \vanact. It is straightforward to verify
this directly along the lines of the calculations presented in appendix B.

The fact that the bulk contribution to the action vanishes for the
Euclidean instanton solutions is a well known phenomenon \ggp\mbgg. The
action comes from a boundary term (see \msmg):
\eqn\actionint{
S_{inst}= - \oint d\Sigma^\mu \partial_\mu \sigma = |Z|_\infty
e^{-\sigma_\infty/2} . 
} 
Here 
\eqn\ccdef{
Z = (L^Iq_I- M_i\tilde q^I)
}
 To obtain this the first equation in
\bpsflob\ was used to express the boundary term 
in terms of the asymptotic values of the  $\sigma$ and $Z$.

There will also be a nontrivial phase dependence  weighting the instanton
amplitudes:
\eqn\phase{
exp(i\theta)= \exp(i \zeta^I q_I +i \tilde\zeta_I\tilde q^I) .
}
The discussion of the phase is not complete without the
investigation of the one loop determinants, but that lies beyond the scope  
of this paper.

The microscopic description of these instantons is well known: they
correspond to M2 branes wrapped on supersymmetric three cycles (special
Lagrangian submanifolds) \bbs. The action is given by the world-volume
action of the wrapped M2 brane on the three cycle in the Calabi-Yau
manifold.  Geometrically the action can therefore be expressed in the
following way
\eqn\geoact{
S_{inst}= e^K|\int_\Gamma \Omega|_\infty+i \int_\Gamma C_\infty ,
}
where $\Gamma$ is the (special Lagrangian) three cycle on which the
membrane instanton is wrapped and the subscript denotes that the fields are
evaluated at asymptotic infinity. The cohomology class of the cycle is
$\Gamma= \tilde q^I\alpha_I- q_I\beta^I$. Using $\int_\Gamma \omega = \int
\omega\wedge \Gamma$ \actionint\ and \phase\ follow from \geoact.

In this section we discussed solutions where $Q_5=0$. If one relaxes this
condition (and equivalently \constrA) the solutions become more
complicated. The microscopic interpretation of such solutions is that of a
'bound state' of M5 and M2 branes wrapped on a CY (in flat space such
configurations were discussed in \lambert). These configurations might also
be related to M5 branes with three form flux turned on
\lustd\minasian. Solutions with $Q_5$ for the universal 
hypermultiplet  were discussed in \msmg. 

In \lusta\ the c-map was used to relate the stationary black hole solutions
of \sabrab\ to instanton solutions in four dimensions. The main
difference 
to the solution in section 9 is that the conditions \fbcvan\ and \constrA\
are relaxed: 
\eqn\relacx{
\eqalign{H_I\partial_\mu \tilde H^I- \tilde H^I\partial_\mu H_I&=
e^{2\sigma}\big( \partial_\mu
\ax+\zeta^I\partial_\mu\tilde\zeta_I-\tilde\zeta^I\partial_\mu\zeta_I\big)\cr
&= {Q_5\over r^4}  .
}}
Even for these solutions the bulk part of the  action of the  instanton
will vanish and the instanton action will be
given by the boundary term \actionint. Using the ansatz for the volume
scalar \volsc\ gives
\eqn\instaqf{
\eqalign{ {1\over 2}\partial_r \sigma&=  {-i\over 2}e^{-\sigma}\big
( \partial_r \bar Z^IF_I+\bar Z^I\partial_r F_I-\partial_M Z^I\bar
F_I-Z^I\partial_r \bar F_I\big)\cr
&= {1\over 2}{1\over r^4} e^{-\sigma/2}\big( \tilde q^I M-q_IL^I+\tilde
q^I\bar M_I- q_I\bar L^I\big)\cr
&= {1\over r^4} e^{-\sigma/2}{Z+\bar Z\over 2}.
}}
For $Q_5\neq 0$ one finds that $Z\neq \bar Z$ in particular
\eqn\zzbare{
{1\over r^4}(Z-\bar Z)= e^{-\sigma} ( H_I\partial_r \tilde H^I-
\tilde H^I\partial_r H_I)= e^{-\sigma}{Q_5\over r^4}.
}
Note that because of the analytic continuation of $\tilde \phi$ in 
Euclidean space one has
\eqn\modzsq{
|Z|^2 = Re(Z)^2 + Im(Z)^2 = Re(Z)^2- Q_5^2
}
and the instanton action becomes
\eqn\instafqfb{
S_{inst}=-\oint d\Sigma^\mu \partial_\mu \sigma= \sqrt{e^{-\sigma}|Z|^2+
e^{-2\sigma}Q_5^2} .
}  
This generalizes the result \actionint\ to the case $Q_5\neq 0$. The
action \instafqfb\ is  characteristic of a non threshold bound state and
has the same form as  the action found in \msmg\ for the universal
hypermultiplet. Note that this is not surprising, since a bound state of
fivebranes and membranes which preserves half the supersymmetry as to be
non-threshold, since  threshold bound state would preserve a quarter of the
supersymmetry.

\newsec{Fermionic Zero Modes}

In addition to the vanishing of the hyperino variation \susyhyp\ the
variation of the gravitino \susygrav\ has to vanish for a BPS
solution. Note that for the solution of section 9  the diagonal elements
of the $Sp(1)$ connection  $Q_A^B$ in \qdeft\ vanish. Using the hyperino
BPS condition \susyvhp\   
the gravitino variation  \susygrav\  becomes
\eqn\susygravb{\delta \psi^{1,2}_\mu= \pm(\partial_\mu \epsilon - {1\over 2}\partial_\mu\sigma)
\epsilon,} 
Hence the supersymmetry parameter takes the form  $\epsilon= e^{\sigma/2}
\epsilon_0$, where 
$\epsilon_0$ is a constant spinor.

The instanton solution has five bosonic zero modes, which  correspond to
collective coordinates  translating the center of the instanton. Since
the instanton is a BPS object in a supersymmetric theory, there are also
fermionic zero modes, corresponding to fermionic collective coordinates.
  The fermionic zero modes can be obtained  by applying the broken
supersymmetries on the bosonic field configuration. There are four 
broken supersymmetries and hence the instanton solutions
will have (at least) four fermionic zero modes.\foot{These are the 'center
of mass' 
zero modes coming from the broken zero modes. In principle there could be
additional zero modes, coming for example from moduli of the supersymmetric
cycle. This is a very interesting question related to calculations of
superpotentials for wrapped branes \kachru\brunner } In a path integral
around the saddlepoint  the
integral over fermionic collective coordinates vanishes 
unless the zero modes are soaked up by field insertions. This leads to
instanton induced interactions a la t'Hooft. In the present case the
instanton will induce four-fermion terms like 
\eqn\fourferm{
\int d^5x\;  \Omega_{IJKL}\;\bar \xi^I\xi^J\; \bar \xi^K\xi^L
e^{-S_{inst}.
}}
The completely symmetric tensor $\Omega_{IJKL}$ is related  to a instanton
modification of the $Sp(n)$ part of the curvature on the quaternionic
geometry of the moduli space.  Supersymmetry relates \fourferm\   to 
corrections to the moduli space metric.

For a complete calculation of such terms one would need a path integral
formulation of M-theory which at the moment does not exist. In particular
the calculation of fluctuation determinants is an important part of the
calculation (see \mooreharvey\ for an eloquent discussion of this issue). 

However the fact that terms like \fourferm\ will be induced and the
'semi-classical' contribution can be determined using that the broken
supersymmetry is given by $\epsilon_1=-\epsilon_2$ and the hyperino
variation  obtained from  the broken supersymmetry is 
\eqn\broksusy{
\delta \xi_1^I = e^{1I}_M\gamma^M\epsilon_1,\quad\quad
\delta\xi_2^I = \bar e_\mu^{1I}\gamma^\mu\epsilon_1 .
}
Using a similar reasoning as for $N=2$ black holes one can use  the BPS
conditions \bpsflob\ to estimate the behavior of the fermionic zero
modes \broksusy\ as $r\to \infty$ and $r\to 0$. Using the fact that
$|Z(r)|\to |Z|_\infty$ as $r\to \infty$ and that the scalar moduli approach
a fixed point as $r\to 0$ one finds 
\eqn\estixi{
\delta \xi^I \sim {1\over r}\quad r\to 0,\quad \quad \delta
\xi^I\sim{1\over r^4},\quad r\to \infty .
}
This implies that the fermionic zero modes will be normalizable and
therefore be will be related to fermionic collective
coordinates. Furthermore the instanton induced fermion four point vertex  
\eqn\fpvert{
\int d^5x_0\; \delta\xi_1^I(x-x_0) \delta\xi^J_2(x-x_0)\delta
\xi_1^K(x-x_0)_1\delta \xi^L_2(x-x_0)
}
is finite, since the integrand will behave as $1/r^4$ as
$r\to 0$ and as $1/r^{16}$ as $r\to \infty$. This integral determines the
'semiclassical' contribution to the four-fermion vertex \fourferm, however
the complicated  
form of the $e^{Ia}_\mu$ makes the closed evaluation of the integral
difficult in general. In type II Calabi-Yau compactifications such
instanton corrections can also be calculated at Gepner points using
conformal field theory and boundary state techniques \yujia\yujib.

\newsec{Relation to Black Holes}

In four dimensions the c-map \cecotti\ relates the special geometry of the
N=2 vector multiplets to the quaternionic geometry of the
hypermultiplets. This can be derived via dimensional reduction  because on a
circle a four dimensional vector field is related to two
scalars by dualization. Hence in the compactified theory a $N=2$
vector multiplet is on-shell equivalent to a $N=2$
hypermultiplet. 

In \lusta\ the c-map was used to relate the most general
stationary BPS black hole solutions found in \sabraa\sabrab\sabrac\ to 
D-instanton  
solutions in four dimensions. It was argued that the stationary black
hole solutions can be reduced along the timelike Killing direction and
a (formal) T-duality on the timelike orbit of this Killing vector
relates the black holes to instantons. 
Since the hypermultiplet geometry is the same
in  four and five dimensions it
is not surprising that the instanton solutions are 
also be solutions in five dimensions, as we showed in this paper.
However the relation to the black hole solutions in five dimensions
dimensions is lost.

Applying the point of view described in section 7 to these solutions
it is clear that the harmonic functions \harmfct\ parameterize a geodesic
null submanifold, where the velocity vectors satisfy constraints
imposed by supersymmetry. 

In the previous section it was shown  that the BPS conditions for
instantons in five
dimensions are isomorphic to the BPS flow equations of $N=2$ black
holes in four dimensions. The instanton action is related to the
asymptotic central 
charge and hence to the ADM mass of the four dimensional black hole,
which depends on the charges and the asymptotic values of the complex
structure moduli (but not on the asymptotic values of
$\zeta^I,\tilde\zeta_I$). In the case of black holes the attractor
mechanism resolves an important puzzle concerning the independence of the
black hole entropy (given by the area of the horizon) from the value of the
asymptotic moduli. It is an interesting question what the analogue of the
area of the horizon and the entropy is for the D-instanton. Note that in
the Einstein frame the instanton solution is flat. In the case of the
ten dimensional D-instanton \gibbons\ it was remarked that 
solution in the string frame has the interpretation of an Euclidean
wormhole. It is quite likely that a similar interpretation of the instanton
solution is possible here and that in the 'string'-frame the instanton
solutions are a Euclidean wormhole and the size of the
neck of the wormhole is related to $Z(q^I,\tilde q_I)|_{fixed}$, which
appeared in the formula for the area of the horizon of the black hole.

\newsec{Gauging}

In the case of instanton solutions of the Euclidean $N=2$ 
supergravity discussed in this paper only the hypermultiplets are
nontrivial. Such solutions 
are saddlepoints of the action and  are responsible for non-perturbative
corrections to the geometry of the hypermultiplet moduli space. 
The solutions are governed
by harmonic functions and can be interpreted as null geodesic submanifolds
in the moduli space manifold. Such an interpretation is only
possible in the Euclidean theory where the moduli space metric components
for the  axion-like scalars 
$\zeta^I,\tilde \zeta_I$ change signature. In the original Minkowskian
spacetime the moduli space metric is positive definite and  the only null
geodesics are given by constant 
maps, i.e. the hypermultiplet solutions are trivial. This implies  that
there are  no
nontrivial hypermultiplet solutions  with all the other fields turned
off. This conclusion not altered  when the 
vector multiplets are taken into account due to the decoupling of vector
and hypermultiplets at the two derivative level demanded by $N=2$
supersymmetry. However if one considers gauged 
$N=2$ supergravity one can obtain nontrivial flows involving the
hypermultiplets. The reason why  nontrivial BPS hypermultiplet dynamics
are possible for gauged supergravity is that the supersymmetry
transformation of the hyperinos is modified \freh\ceres\ :
\eqn\hypegag{
\delta \zeta^{A\alpha} = V_u^{A\alpha}
(\gamma^\mu \partial_\mu 
\phi^u+ iX^Ik_I^u)\epsilon .
} 
Here $A$ and $\alpha$  are  the $Sp(1)$ and $Sp(n_h)$ index of the hyperino 
respectively and $\phi^u$ denotes the scalars in the hypermultiplets and
$V$ is the quaternionic vielbein which can be read of from \vielba. 
Here  $X^I$ denotes the real
scalars of the five dimensional (very) special geometry of the
vector multiplets. 

The new term in the supersymmetry transformation
contains the 
Killing vector fields $k^I$ which generates $n_v+1$ isometries of the
quaternionic manifold $\phi^u\to \phi^u +k_I^u\epsilon^I$. The BPS condition
following from  \hypegag\ can be 
nontrivial in Minkowski space and for example allows for domain wall
solutions where there are nontrivial flow equations governing the evolution
of the hypermultiplets \gukov\louisb. The non-perturbative corrections to
the hypermultiplet metrics can also be important since instanton effects
might destroy isometries of the hypermultiplet moduli space which cannot be
gauged. It would be very interesting to investigate this further.

\newsec{Conclusions}

In this paper Euclidean instanton solutions of  $N=2$ supergravity were
found for theories with arbitrary number of hypermultiplets. The BPS
conditions are isomorphic to the attractor equations for $N=2,d=4$ black
holes.  The instanton actions are basically given by the volume of the 
three cycle on which the membranes are wrapped. In the case of non
vanishing fivebrane 
charge the action has a form reminiscent of a non threshold bound
state. The investigations reported here were limited to supergravity: both 
the inclusion 
of higher derivative corrections like $R^4$ \vgg\ terms in the dimensional
reduction of eleven dimensional supergravity along the lines of
\antond\stromingera\ would be very interesting. On a microscopic level it
would be interesting to understand better the question of extra moduli
associated with Euclidean wrapped membranes, if such extra moduli are not
lifted by higher terms in the instanton action (which are analogous to
superpotentials in spacetime filling branes \kachru) such instantons would
not contribute to corrections to hypermultiplet geometries. The best hope
one might have to calculate instanton corrections is by heterotic-type II
duality. However even though on the heterotic side the hypermultiplets
geometry is  determined at tree level the situation is complicated due to
world sheet instanton corrections (see \aspinwall\ for a discussion of
these issues).  The form of the instanton action for non vanishing fivebrane
charge \instafqfb\ generalizes the one found for the universal
hypermultiplet \msmg\ to any CY compactification, it would be very
interesting to  analyze this from the perspective of the world-volume theory
of the fivebrane wrapped on the CY.

\medskip
{\bf Acknowledgements}
\medskip

We are grateful to A. Strominger for very useful conversations.
The work of M.G. is supported in part by the David and Lucile Packard
Foundation. The work of M.S. was supported in part by NSF grants 
DMS-97-09694 and PHY-98-02709.

\appendix{A} {Special Geometry}

A canonical cohomology basis  $(\alpha_J,\beta^J),\; J=0,\cdots,h_{2,1}$ 
of three forms satisfies 
\eqn\dualcom{
\eqalign{&\int\alpha_I\wedge\alpha_J=0,\quad \int
\beta^I\wedge\beta^J=0 , \cr
& \int \alpha_I\wedge \beta^J=-\int \beta^J\wedge\alpha_I=\delta_I^J .
}}
The periods over $A^I,B_I$ cycles dual to the cohomology basis define
projective coordinates of the moduli space of complex structure
deformations:
\eqn\compst{
Z^I=\int_{A^I}\Omega,\quad  F_I = \int_{B_I}\Omega ,
}
where $F_I = \partial_I F$  and (if it exists) the prepotential $F$ is a 
homogeneous polynomial in $Z^I$ 
of weight two. The holomorphic three form can be expressed in the
canonical cohomology basis in the following way
\eqn\canbasis{
\Omega = Z^I\alpha_I- F_I \beta^I.
} 
The metric for the  the scalars $z^k$ is determined by the  K\"ahler
potential 
\eqn\kaehler{
K= -\ln\big(i(\bar Z^IF_I-Z^I\bar F_I)\big).
}
A variation of the holomorphic three form $\Omega$ is given
\eqn\varthree{
\partial_I \Omega =\Omega_I= -K_I\Omega +\Phi_I
}
where $\Phi_I$ is a $(2,1)$ form and  
\eqn\kdefin{
K_I= -{N_{IJ}\bar Z^J\over \bar Z^K N_{KL}Z^L}
}
and 
\eqn\capndef{
N_{IJ} = (Im\;F)_{IJ}
}

It is convenient to define the covariantly constant sections $L^I=
e^{K/2}Z^I, M_I=e^{K/2}F_I$ which satisfy $D_{\bar i} L^I= (\partial_{\bar
i} -\half\partial_{\bar i}K)L^I=0$. The following symplectic sections
are defined as
\eqn\spdefl{\eqalign{
f_i^I &= D_i L^I=e^{K/2}(\partial_i+\partial_i K)Z^I, \cr
h_{iI}&=D_i M_I =e^{K/2}(\partial_i+\partial_i K)F_I.
}}
which satisfy the following equations
\eqn\spgeomb{
D_{\bar j}f_i^I= g_{i\bar j}L^I,\quad  D_{\bar j}h_{iI}=
g_{i\bar j}M^I, \quad D_i f_j^I=
iC_{ijk}g^{k\bar l} \bar f_{\bar l}^I, \quad D_i h_{jI}=
iC_{ijk}g^{k\bar l} \bar h_{\bar lI} 
}
Note that $D_i  f_j^I= D_j f_i^I$, which follows from the total symmetry of
$C_{ijk}$.

To carry out the dimensional reduction procedure one needs the
following relation expressing the property of the cohomology basis under
Hodge duality  
\eqn\hodged{
\eqalign{
\star\alpha_I&= -\big(Im({\cal N}) + Re( {\cal N})
Im({\cal  N})^{-1}Re( {\cal N})\big)_{IJ}\beta^J+Re( {\cal N}) Im({\cal 
N})^{-1}_I\;^J\alpha_J, \cr
\star\beta^I&= Im({\cal N})^{-1I}_{J}\alpha^J- Im({\cal
N})^{-1\;IJ}Re( {\cal N})_{JK}\beta^K,
}}
where
\eqn\nmatdef{
{\cal N}_{IJ} =\bar F_{IJ}+2i{N_{IK}Z^K N_{JL}Z^L\over 
Z^P N_{PQ} Z^Q} .
}
This matrix satisfies the important relations
\eqn\nlm{
{\cal N}_{IJ}L^J = M_I 
}
and
\eqn\hnf{
h_{iI}=\bar{\cal N}_{IJ} f^J_i .
}

\appendix{B}{Supersymmetry Transformations from Eleven Dimensions}

In this appendix we indicate how to derive the supersymmetry transformation 
rules \susyhyp\ for the hyperinos from the eleven dimensional ones
\susytraf\ by dimensional reduction.
On a Calabi-Yau threefold  there are two covariantly constant spinors which
are related to the existence of Killing spinors of unbroken N=2 spacetime
supersymmetry. Using the commutation relation $\{\gamma_i,\gamma_{\bar
j}\}=2g_{i\bar j}$ one can define the two spinors by demanding
\eqn\spindf{
\gamma_i \mid \Omega\rangle=0\;,i=1,2,3\;\quad \quad\quad
\gamma_{\bar 
i}\mid \bar \Omega \rangle=0,\; \bar i=1,2,3
}
where $\mid \Omega\rangle = 1/|\Omega|^2 \Omega^{ijk}\gamma_{ijk}\mid \bar 
\Omega\rangle$. 
Decomposing the eleven dimensional gamma matrices as $\Gamma_\mu=
\gamma_\mu\otimes \gamma_7, \mu = 0,\cdots 3$ and $\Gamma_i = 1\otimes
\gamma_i, i=1,\cdots 6$. One can expand the Killing spinor which
parameterizes the unbroken $N=2$ supersymmetry as $\eta = \epsilon_1\otimes
\mid \Omega\rangle+ \epsilon_2\otimes \mid \bar \Omega\rangle$ where
$\epsilon_{1,2}$ are five dimensional (symplectically real) spinors.
Massless fermions in the (non universal) hypermultiplets can be
constructed using the 
covariantly constant spinors and harmonic $H^{2,1}$ and $H^{1,2}$
forms. The zero modes of the Dirac operator on the CY-manifold,
transform as $3$ and $\bar 3$ under the $SU(3)$ holonomy respectively 
\eqn\covarsp{
\psi_i^{(k)}= h^{(k)}_{i\bar j\bar k} \gamma^{\bar
j\bar k}\mid \bar \Omega \rangle,\quad\quad \psi_{\bar i}^{(k)}=
h^{(k)}_{\bar i jk} \gamma^{ j k}\mid \Omega \rangle .
}  
Here $h^{(k)}_{i\bar j\bar l}$ and $h^{(k)}_{\bar i jl}$ are
the $H^{1,2}$ and $H^{2,1}$ forms  and $\psi_i^{(k)}, \psi_{\bar
i}^{(k)}$ are identified with the hyperinos
$\xi_1^{k},\xi_2^{k},k=1.\cdots,h_{2,1}$. 
 
An arbitrary three form $\Gamma$ can be decomposed in terms of the elements
of  $H^{3,0}\oplus H^{2,1}\oplus H^{1,2}\oplus H^{0,3}$ \denef.
\eqn\decompa{
\Gamma= i \bar Z(\Gamma)\Omega - ig^{i\bar j}\bar D_{\bar
j}\bar Z(\Gamma)D_i
\Omega +ig^{ i \bar j} D_{
i} Z(\Gamma)D_{\bar j}
\Omega - i  Z(\Gamma)\bar \Omega ,
}
where $Z(\Gamma)= e^{K/2}\int \Gamma\wedge \Omega$. For the field strength 
associated with the potential \aexans\ $F_\mu=
\partial_\mu \zeta^I \alpha_I +\partial_\mu \tilde \zeta_I \beta^I$ gives
\eqn\decomb{
F_\mu = i \bar L^I(\bar {\cal
N}_{IJ}\partial_\mu\zeta^J+\partial_\mu\tilde\zeta)\Omega
-i g^{i \bar j}\bar f_{\bar
j}^I ({\cal
N}_{IJ}\partial_\mu \zeta^J+\partial_\mu \tilde\zeta_I)
D_i\Omega+c.c.
}
The scalars $z^i$ are the moduli associated with complex structure
defomation of the Calabi-Yau manifold. The spin connection
$\omega^{\bar i}_{\mu j},\omega^{ i}_{\mu\bar j} $ can be expressed in
terms of the Christoffel connections 
\eqn\expandsc{\Gamma^{\bar j}_{\mu i}= {1\over 2 |\Omega|^2}\Omega^{\bar
j\bar l \bar k} \bar h^{(i)}_{\bar l\bar k j}\partial_\mu \bar z^i,\quad
\Gamma^{ j}_{\mu \bar i}= {1\over 2 |\Omega|^2}\Omega^{
j l  k} h^{(i)}_{ l k \bar j}\partial_\mu  z^i}
by appropriate multiplication with vielbeins.
Here $|\Omega|^2=1/3! \Omega_{ijk}\bar \Omega^{ijk}$ and $h^{(k)}_{ ij
\bar l }$ are the componets of the (2,1) form $D_k \Omega$. 
Plugging the formulas for the spin connection and the gauge field
given in this appendix into the supersymmetry
transformation \susytraf\ of the eleven dimensional gravitino one can
derive the form of the five dimensional supersymmetry transformation
\vielba\ where the vielbein $e^I$ with $I=0$ is associated with $\Omega$
and $e^k$ $k=1,\cdots,h_{2,1}$ with $D_k\Omega$ respectively.

\appendix{C}{Equations of Motion}

In order to show that the ansatz satisfies the $\sigma$ equation of motion
\eqofmhb\ one calculates
\eqn\sigmem{
\eqalign{{1\over 2}
\nabla^2 \sigma &={1\over 2}\big( {d^2 \sigma\over dr^2}+ {4\over
r}{d \sigma\over dr}\big)\cr 
&= -(L^Iq_I-M_I\tilde q^I)^2 {e^{-\sigma}\over r^8}- {d\over
dr}(L^Iq_I-M_I\tilde q^I) {e^{-\sigma/2}\over r^4}\cr
&= -\Big((L^Iq_I-M_I\tilde q^I)^2-(f_{i}^Iq_I- h_{iI}\tilde q^I) g^{i\bar j}
(\bar f_{\bar i}^Iq_I- \bar h_{\bar iI}\tilde q^I)\Big) {e^{-\sigma}\over
r^8}\cr
&= {1\over 2}e^{\sigma} ({\cal N}_{IK} \partial_\mu \zeta^K +\partial_\mu \tilde
\zeta_I)(Im 
{\cal N}^{-1})^{IJ}(\bar {\cal N}_{JL} \partial_\mu \zeta^L +\partial_\mu
\tilde \zeta_J)}
}
(where repeated use has been made of \bpsflob\ ).
To show that \eqofmc\ is satisfied it is convenient to start from
\eqn\zeofmd{\eqalign{\partial_\mu (g_{\bar k l}\partial^\mu z^{ l}) =
\partial_{\bar i}(\bar f_{\bar k}^Iq_I- \bar h_{\bar kI}\tilde q^I) g^{\bar
  i l} (f_{l}^Iq_I- h_{lI}\tilde q^I){e^{-\sigma}\over r^8}\cr
+ \partial_{ i}(\bar f_{\bar k}^Iq_I- \bar h_{\bar kI}\tilde q^I) g^{
i \bar l} (\bar f_{\bar l}^Iq_I- \bar h_{\bar l I}\tilde
q^I){e^{-\sigma}\over r^8} +(\bar f_{\bar k}^Iq_I- \bar h_{\bar kI}\tilde
q^I)(L^Iq_I-M_I\tilde q^I){e^{-\sigma}\over r^8}}}
together with 
\eqn\zeofme{
(\partial_{\bar i} g_{\bar k l})\partial_\mu z^{\bar i}\partial_\mu z^l= -
\partial_{\bar i}g^{\bar k l} (f_{l}^Iq_I- h_{lI}\tilde q^I)(\bar f_{\bar
k}^Iq_I- \bar h_{\bar kI}\tilde q^I){e^{-\sigma}\over r^8.}
}
Using the fact that $\partial_k K \partial_\mu z^k- \partial_{\bar k}K
\partial_\mu z^{\bar k}=0$ and several other identities given in
apendix B one arrives at
\eqn\reseqm{\eqalign{
\partial_\mu (g_{\bar k l}\partial_\mu
z^{l})-(\partial_{\bar m}  
g_{\bar k l})\partial_\mu z^{\bar m}\partial_\mu z^l&=(\bar f_{\bar
k}^I q_I- \bar h_{\bar k I}\tilde q^I)(L^Iq_I-M_I\tilde
q^I){e^{-\sigma}\over r^8}\cr 
&+\partial_{\bar k}\Big((f_{l}^Iq_I- h_{lI}\tilde q^I) g^{l\bar m}
(\bar f_{\bar m}^Iq_I- \bar h_{\bar m I}\tilde q^I)\Big) {e^{-\sigma}\over 
r^8}.   
}}
The other term in the equation of motion \eqofmc\ is given by
\eqn\zeofmf{\eqalign{&{1\over
2}e^{\sigma} 
 \partial_{\bar k} \Big\{ Im{\cal
N}_{IJ}\partial_\mu 
    \zeta^I\partial^\mu\zeta^J
+(Im{\cal N}^{-1})^{IJ}\big(Re{\cal
N}_{IK}\partial_\mu\zeta^K+\partial_\mu\tilde\zeta_I\big)\big(Re{\cal
N}_{JL}\partial_\mu\zeta_L+\partial_\mu\tilde\zeta_J\big)\Big\}\cr
&= {1\over 2} {e^{-\sigma}\over r^8} \partial_{\bar k}\Big\{ (q_I-{\cal
    N}_{IJ}\tilde q^J)(Im{\cal N}^{-1})^{IK}(q_K-\bar{\cal
    N}_{KL}\tilde q^L)\Big\}\cr
&=- {e^{-\sigma}\over r^8} \partial_{\bar k}\Big\{(f_{l}^Iq_I-
 h_{lI}\tilde q^I) g^{l\bar m} 
(\bar f_{\bar m}^Iq_I- \bar h_{\bar mI}\tilde q^I)+L^I(q_I-{\cal
    N}_{IJ}\tilde q^J)\bar L^K (q_K-\bar{\cal
    N}_{KL}\tilde q^L)\Big\} ,
}}
where in the last line the identity \identuse\ was used. The last term in 
the third line of \zeofmf\ can be rewritten as follows
\eqn\rewrLL{\eqalign{&\partial_{\bar k}\Big\{L^I(q_I-{\cal
    N}_{IJ}\tilde q^J)\bar L^K (q_K-\bar{\cal
    N}_{KL}\tilde q^L)\Big\}\cr
&= D_{\bar k}\big(L^I(q_I-{\cal
    N}_{IJ}\tilde q^J)\big)\bar L^K (q_K-\bar{\cal
    N}_{KL}\tilde q^L)+L^I(q_I-{\cal
    N}_{IJ}\tilde q^J)D_{\bar k}\big(\bar L^K (q_K-\bar{\cal
    N}_{KL}\tilde q^L)\big)\cr
&=(L^Iq_I-M_I\tilde q^I)(\bar f_{\bar k}^Kq_K- \bar h_{\bar kK}\tilde
    q^K). }}
Adding \reseqm\ and \zeofmf\ und using \rewrLL\  equation \eqofmc\ follows.

\listrefs

\end